\documentclass[
    ,final            
  ]    {aipproc}
\layoutstyle{8x11single}

\def\prd#1#2#3{{\it Phys.\ Rev.} {\bf D#1}, #2 (20#3)}
\def\prl#1#2#3{{\it Phys.\ Rev.\ Lett.} {\bf #1}, #2 (20#3)}
\def\etal {{\it et al.}}

\newcommand{\ra}{\rightarrow}
\newcommand{\Eq}[1]{Eq.~(\ref{eq#1})}
\newcommand{\beq}{\begin{equation}}
\newcommand{\eeq}{\end{equation}}

\newcommand{\num}{\nu_\mu}
\newcommand{\nue}{\nu_e}
\newcommand{\nmb}{\bar{\nu}_\mu}
\newcommand{\neb}{\bar{\nu}_e}
\newcommand{\meb}{\nmb \ra \neb}
\newcommand{\sint}[1]{\sin^2 2\theta_{\mu #1}}
\newcommand{\pnot}{P(\nmb \ra \mathrm{Not}\; \nmb) }

\newcommand{\gtwid}{\mathrel{\raise.3ex\hbox{$>$\kern-.75em\lower1ex\hbox{$\sim$}}}}
\newcommand{\ltwid}{\mathrel{\raise.3ex\hbox{$<$\kern-.75em\lower1ex\hbox{$\sim$}}}}

\begin{document}

\title{Are There Sterile Neutrinos?
	\footnote{FERMILAB-CONF-14-023-T. To appear in the Proceedings of the CETUP Workshop on Neutrino Physics and Astrophysics.} }

\classification{14.60.St}			
\keywords      {Sterile neutrinos}

\author{Boris Kayser}{address={Fermi National Accelerator Laboratory \\ P.O. Box 500, Batavia IL 60510} }

\begin{abstract}
We update the hints of the existence of sterile neutrinos.
\end{abstract}

\maketitle



The hints that there may be sterile neutrinos continue to be intriguing, and to call for experiments that hopefully will tell us definitively whether such neutrinos exist or not. In ``Tensions With the Three-Neutrino Paradigm'', \cite{ref1} we described the hints of sterile neutrinos, and some of the ideas for future experiments to probe whether they exist, as of June, 2012. Here we update the picture.

	The MiniBooNE experiment has reported new results \cite{ref2} that show evidence for both $\meb$ and  $\num \ra \nue$ oscillation at $L/E \sim 1$\, m/MeV, where $L$ is the distance travelled by the neutrinos between creation and detection, and $E$ is their energy. These results, like the earlier evidence for $\meb$ from the LSND experiment, \cite{ref3} suggest the existence of at least one neutrino squared-mass splitting $\Delta m^2$ larger than approximately 0.1 eV$^2$. If there is only one such large splitting, then at short baselines the probability for $\meb$, $P(\meb)$, is given by
\beq
P(\meb) \simeq \sint{e} \, \sin^2 \left[ 1.27 \Delta m^2 (\mathrm{eV}^2 ) \frac{L \mathrm{(m)}} {E \mathrm{(MeV)}} \right] ~~.
\label{eq1}
\eeq
Here,  $\sint{e}$ is a mixing parameter that satisfies $0 \leq \sint{e} \leq 1$ . If one separately fits the LSND antineutrino data and the MiniBooNE antineutrino data with \Eq{1}, one finds considerable overlap between the  $(\Delta m^2 - \sint{e})$ regions allowed by each of the data sets. \cite{ref2}

The hints of sterile neutrinos include the ``reactor anomaly''. This is the observation that the reactor $\neb$  flux measured by detectors that are only (10 -- 100) m from reactor cores is $\sim$ 6\% below the theoretically expected value. \cite{ref4}  If the missing flux has disappeared by oscillating into another flavor or flavors, this behavior, like that observed in LSND and MiniBooNE, points to a splitting $\Delta m^2$ larger than $\sim$ 0.1 eV$^2$. A recent analysis finds that if one uses the now-known value of the mixing angle $\theta_{13}$, and takes into account $\neb$ flux measurements at detectors that are about 1 km from their reactors, the missing flux is reduced to $\sim$ 4\% of the theoretically expected value, and the significance of the discrepancy is reduced to 1.4\,$\sigma$. \cite{ref5} The story of the reactor anomaly no doubt will continue. It would be desirable to see if  $\neb$ flux that is produced by a radioactive source, rather than a reactor, disappears as well. 

While  $\neb$   may disappear while traveling a short distance, there is no evidence that  $\nmb$  do so. If there is only one large $\Delta m^2$, then at short baselines the probability of $\nmb$ disappearance,  $\pnot$, is given by  
\beq
\pnot  \simeq  \sint{\mu} \, \sin^2 \left[ 1.27 \Delta m^2 (\mathrm{eV}^2 ) \frac{L \mathrm{(m)}} {E \mathrm{(MeV)}} \right] ~~.
\label{eq2}
\eeq
Here, $\sint{\mu}$  is a mixing parameter that is distinct from $\sint{e}$ and satisfies $\sint{e} \leq \sint{\mu} \leq 1$. A joint analysis of data from the SciBooNE and MiniBooNE detectors, which were both in the same beamline at Fermilab, with SciBooNE closer than MiniBooNE to the antineutrino source, has excluded values of $\sint{\mu}$ greater than a limit that ranges from unity down to  10$^{-2}$, depending on the value of $\Delta m^2$. \cite{ref6}

Several interesting constraints relate disappearance probabilities to appearance probabilities. Assuming only CPT invariance, we must have $P(\neb \ra \mathrm{Not}\; \neb) \geq  P(\meb)$. The appearance probability on the right-hand side of this constraint is reported to be 0.0026 by LSND. \cite{ref3} If the disappearance probability on the left-hand side is $\sim$ 0.06, as suggested by the reactor data, the constraint is very comfortably satisfied.  Now let us assume that we have a ``3 + 1'' spectrum of neutrino mass eigenstates; that is, the three states $\nu_{1,2,3}$ of the standard three-neutrino paradigm, plus a largely-sterile fourth state $\nu_4$ separated from $\nu_{1,2,3}$ by a splitting $\Delta m^2$ that is larger than $\sim$ 0.1 eV$^2$. Then, as is easily shown, experiments whose $L/E$ is too small for them to be sensitive to the small splittings among $\nu_{1,2,3}$, but large enough so that they average over the rapid oscillation driven by the large splitting between $\nu_{1,2,3}$ and $\nu_4$, should find that $\pnot P(\neb \ra \mathrm{Not}\; \neb) \simeq 2P(\meb)$. Finally, let us assume that we have a ``3 + 2'' spectrum, with two largely-sterile mass eigenstates $\nu_4$ and $\nu_5$ in addition to the neutrinos $\nu_{1,2,3}$ of the three-neutrino picture. Then experiments that are insensitive to the splittings among $\nu_{1,2,3}$ but average over the rapid oscillations driven by the large splitting between  $\nu_{1,2,3}$ and $\nu_4$  and that between $\nu_{1,2,3}$  and $\nu_5$  should find that $\pnot P(\neb \ra \mathrm{Not}\; \neb) \gtwid 2P(\meb)$. \cite{ref7} This constraint and the very similar one for the 3 + 1 spectrum suggest that $\pnot$ and $P(\neb \ra \mathrm{Not}\; \neb)$ should each be of the order of the square root of $2P(\meb)$, which is to say the square root of $\sim$ 0.005. Sensitivity to at least this degree of short-baseline disappearance is a goal of future probes. 

The Planck satellite has now provided very interesting cosmological information \cite{ref8} that bears on the question of whether sterile neutrinos exist. Planck reports the effective number of relativistic degrees of freedom, $N_\mathit{eff}$, at an early time probed by the CMB. The quantity $N_\mathit{eff}$ is so defined that together the three active neutrinos $\nu_{e, \mu, \tau}$ lead to $N_\mathit{eff}$  = 3.046. If, at the time probed, a sterile neutrino is fully relativistic and, as expected from the degree of mixing with active neutrinos required by the terrestrial hints, is present in the number per unit volume that corresponds to statistical equilibrium, then this neutrino will contribute one unit to $N_\mathit{eff}$. Now, the Planck results include the value $H_0 = 67.3 \pm 1.2$ (km/sec)/Mpc for the Hubble constant. There is some disagreement between this value and the value $H_0 = 73.8 \pm 2.4$ (km/sec)/Mpc obtained by other means. \cite{ref9} If Planck does not take this other value of $H_0$ into account, it finds the most probable value of $N_\mathit{eff}$ to be approximately 3.3, but if it does take this other value of $H_0$  into account, it finds the most probable value to be approximately 3.6. Curves showing the probabilities of various possible values of $N_\mathit{eff}$ would appear to allow this parameter to be 3 and also to allow it to be 4, leaving the question of whether there is a sterile neutrino flavor in addition to the three active flavors open. 

Planck also reports that, assuming that in the early universe the number density of every neutrino mass eigenstate is as required by statistical equilibrium, the sum of the masses of {\em all} the mass eigenstates is less than 0.23 eV. Obviously, there is some tension between this bound and any terrestrially-observed splitting $\Delta m^2$ that is larger than 0.1 eV$^2$. However, the bound is subject to parameter degeneracies, and does rest, as mentioned, on the assumption of statistical-equilibrium number densities. 

Thorough new efforts \cite{ref10,ref11} have been made to create global fits to the terrestrial short-baseline neutrino oscillation data and other data relevant to the question of the existence of sterile neutrinos. The spectra that are tried include 3 + 1, 3 + 2, 1 + 3 + 1 (i.e., one extra mass eigenstate lighter, and one heavier, than $\nu_{1,2,3}$), and 3 + 3. In these attempted fits, there is tension between different data, notably between appearance and disappearance data. This has led to speculation that the apparent $\nue$  and $\neb$ excesses seen at the lowest energies probed by the MiniBooNE $\num \ra \nue$ and $\meb$ appearance searches may not be due to neutrino oscillation. Some evidence that the low-energy $\nu_e$ excess may not be due to oscillation has come from the ICARUS and OPERA experiments. These experiments have searched for $\num \ra \nue$ at $L/E \sim 35$ m/MeV, an $L/E$  larger than those of LSND and MiniBooNE, but such that one is still not very sensitive to oscillations driven by the squared-mass splittings among $\nu_{1,2,3}$, while being sensitive to those driven by splittings larger than $\sim$ 0.03 eV$^2$. From negative results, ICARUS  and OPERA disfavor a $\num \ra \nue$ origin of the low-energy $\nu_e$ excess reported by MiniBooNE, \cite{ref12,ref13} although the strength of this disfavoring has been questioned. \cite{ref14} It has been argued that if  the low-energy MiniBooNE data are excluded, then the hypothesis of oscillation driven by a 4-neutrino 3 + 1 spectrum fits the remaining data, and provides a description that is not significantly less good than that based on a  5-neutrino spectrum. \cite{ref15}

There are now many creative ideas for future experiments that hopefully will definitively determine whether sterile neutrinos exist or not. Some of these ideas were described in Ref.~\cite{ref1}. Others are described in a recent Fermilab academic lecture by David Schmitz. \cite{ref16} The ideas include probes of both appearance and disappearance, and involve experiments with neutrinos or antineutrinos from radioactive-decay sources, reactors, accelerators, and a muon storage ring. Confirmation of the existence of sterile neutrinos would have far-reaching implications. Hopefully, the future experiments will either conclusively confirm their existence, or firmly rule them out. 

\begin{theacknowledgments}
  It is a pleasure to thank Barbara Szczerbinska, Baha Balantekin, and Kaladi Babu for the generous hospitality of the CETUP workshop on neutrino physics and astrophysics, and for creating a very effective and enjoyable workshop.
\end{theacknowledgments}

\bibliographystyle{aipproc}

\begin{thebibliography}{29}

\bibitem{ref1}
B. Kayser, \emph{2012 Electroweak Interactions and Unified Theories},
 eds. E. Aug\'{e}, J. Dumarchez, J.-M. Fr\`{e}re, L. Iconomidou-Fayard, and J. Tr\^{a}n Thanh V\^{a}n (ARISF, 2012), p. 191. arXiv: 1207.2167.

\bibitem{ref2}
MiniBooNE Collaboration (A. Aguilar-Arevalo \etal), \prl{110}{161801}{13}. arXiv: 1303.2588.

\bibitem{ref3}
LSND Collaboration  (A. Aguilar-Arevalo {\it et al.}), \prd{64}{112007}{01}. e-Print: hep-ex/0104049.

\bibitem{ref4}
G. Mention \etal, \prd{83}{073006}{11}. arXiv: 1101.2755.

 \bibitem{ref5}
 C. Zhang, X. Qian, and P. Vogel, \prd{87}{073018}{13}. arXiv: 1303.0900.
 
 \bibitem{ref6}
 MiniBooNE and SciBooNE Collaborations (G. Cheng \etal), \prd{86}{052009}{12}. arXiv: 1208.0322.
 
 \bibitem{ref7}
 This relation was derived by J. Conrad, B. Kayser, and J. Kopp. See also M. Maltoni and T. Schwetz, \prd{76}{093005}{07}. arXiv: 0705.0107.
 
 \bibitem{ref8}
 Planck Collaboration (P. Ade \etal), arXiv: 1303.5076.
 
 \bibitem{ref9}
 A. Riess \etal, {\it Astrophys. J.} {\bf 730}, 119, (2011). arXiv: 1103.2976.
 
 \bibitem{ref10}
 J. Conrad, C. Ignarra, G. Karagiorgi, M. Shaevitz, and J. Spitz, {\em Adv. High Energy Phys.} {\bf 2013}, 163897 (2013). arXiv: 1207.4765.
 
 \bibitem{ref11}
 J. Kopp, P. Machado, M. Maltoni, and T. Schwetz, {\em JHEP} {\bf 1305}, 050 (2013). arXiv: 1303.3011.
 
 \bibitem{ref12}
 M. Antonello \etal, {\em Eur. Phys. J.} {\bf C73}, 2345 (2013). arXiv: 1209.0122.
 
 \bibitem{ref13}
 OPERA Collaboration (N. Agafonova \etal), \emph{JHEP} {\bf 1307}, 004 (2013). arXiv: 1303.3953.
 
 \bibitem{ref14}
 W. Louis, private communication.
 
 \bibitem{ref15}
 C. Giunti, M. Laveder, Y. Li, and H. Long, \prd{88}{073008}{13}. arXiv: 1308.5288.
 
 \bibitem{ref16}
 The slides, some with references, that were used in this lecture are available at 
 https://indico.fnal.gov/conferenceDisplay.py?confid=7309.

\end{thebibliography}

\end{document}